# Dopamine modulation of prefrontal delay activity - Reverberatory activity and sharpness of tuning curves

Gabriele Scheler+ and Jean-Marc Fellous*
+Sloan Center for Theoretical Neurobiology
*Computational Neurobiology Laboratory
The Salk Institute
10100 N. Torrey Pines Road - La Jolla, CA 92037

Recent electrophysiological experiments have shown that dopamine (D1) modulation of pyramidal cells in prefrontal cortex reduces spike frequency adaptation and enhances NMDAtransmission. Using four models, from multicompartmental to integrate and fire, we examine the effects of these modulations on sustained (delay) activity in a reverberatory network. We find that D1 modulation may enable robust network bistability yielding selective reverberation among cells that code for a particular item or location. We further show that the tuning curve of such cells is sharpened, and that signal-to-noise ratio is increased. We postulate that D1 modulation affects the tuning of "memory fields" and yield efficient distributed dynamic representations.

## *Introduction*

Recordings from the dorsolateral prefrontal cortex of behaving monkeys on delayed visuospatial tasks, have shown enhanced neuronal firing during the delay period in a subset of neurons that represent the specific object location to be memorized while other cells, coding for different spatial locations were slightly inhibited [6,11]. This delay activity, as well as the overall performance on this task, but not on a simpler sensorimotor task, was decreased by local infusion of
D1 receptor antagonists [10].
There have been several computational models of delay activity in prefrontal cortex. Most use a reverberatory mode of activity that relies on specific predefined synaptic patterns to achieve a stable state of firing activity [1,2,18]. Because of the transient nature of working memory, it is however likely that attractor-type synaptic weight patterns might not have the time to form, in order to store a specific memory item. An alternative view has recently been proposed that involves network bistability, through NMDA receptor activation [5,15]. According to this view, specific patterns of active cells can be dynamically created and reset in few hundreds milliseconds, while synaptic weights remain unchanged.
We present here three models from single cell to large network that illustrate this idea. In a fourth model we show that this mechanism contributes to the sharpening of the tuning curves of an attractor-based network that stores 5 features.

## *Results*

In a multicompartmental model of a reconstructed pyramidal cell, we have shown that the voltage-dependence of NMDA channels was sufficient to induce network bistability [5]. The model neuron received 150 AMPA/NMDA and 30 GABAA presynaptic Poisson distributed synaptic inputs. Excitatory inputs were uniformly distributed on the dendritic tree (1 Hz average individual discharge), while inhibitory inputs were perisomatic (5 Hz average individual discharge).Intrinsic currents included INa, IK, IAHP, ICa, and a calcium pump in the soma, and INa, INap, IM in the dendrites (kinetics were tuned for 36 oC, and are available upon request). In this model, the synaptic inputs were not sufficient to induce spiking when the cell was initially at rest (about -70 mV). However, a brief somatic depolarization was sufficient to depolarize the distal dendrites to the point were the incoming NMDA current was strong enough to maintain the firing of the cell for several seconds after the somatic depolarization was terminated (Fig 1A).

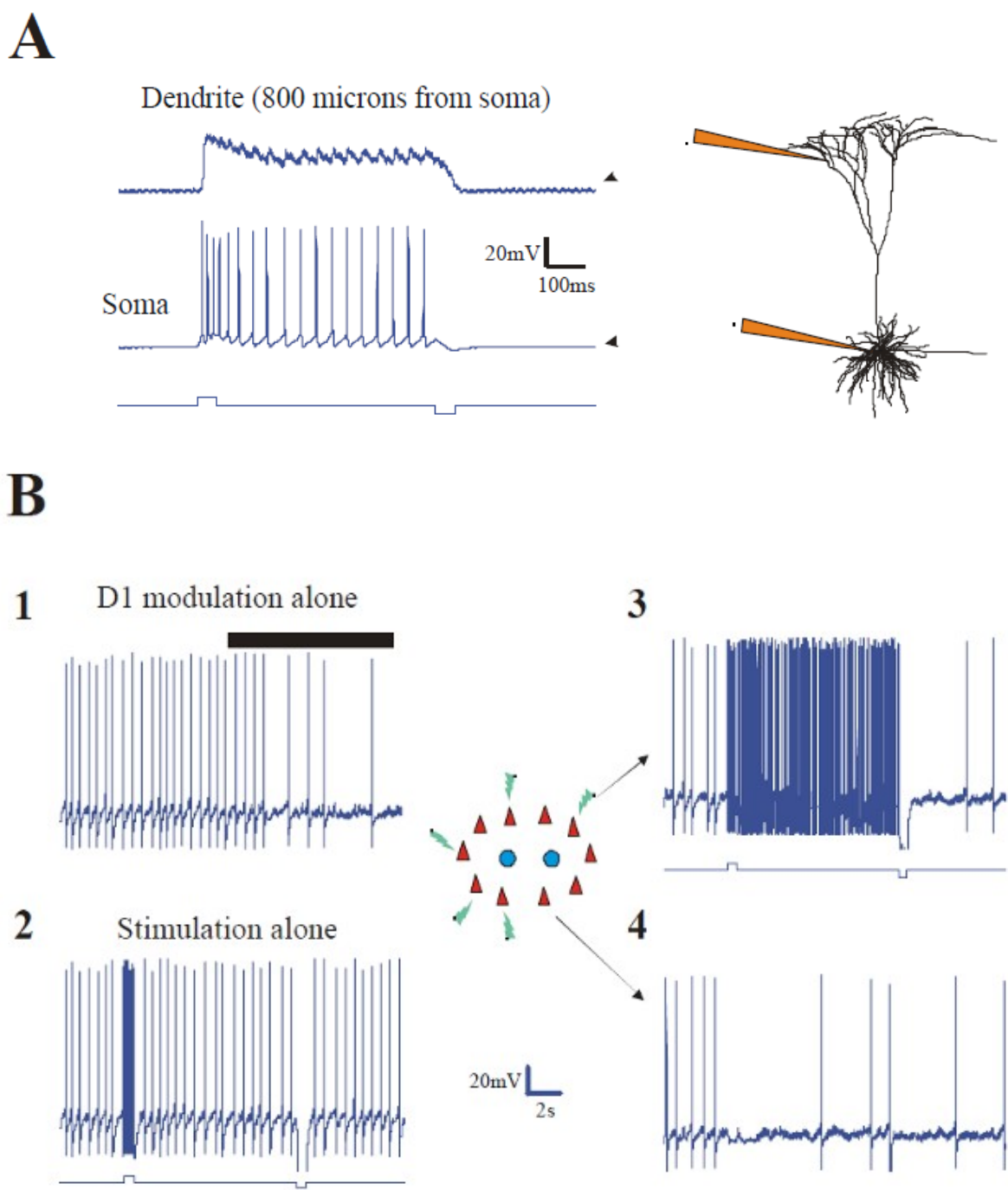

**Figure 1: Multi compartmental models of NMDA induced network bistability. A:** Reconstructed cell bombarded by AMPA/NMDA and GABAA synaptic inputs. A brief somatic current injection is sufficient to depolarize the dendrites of this cell, so that the incoming NMDA current maintains enough overall depolarization so that the cell keeps on firing. Firing is stopped by a brief hyperpolarization. **B:** Fully connected network of simplified multi-compartment cells including 10 pyramidal cells, and 2 interneurons. All synaptic weights between pyramidal cells are equal. 1) The introduction of D1 modulation reduces the spontaneous firing rates of pyramidal cells, and increases the spontaneous firing rate of the interneurons (not shown). 2) Without D1 modulation, a brief somatic depolarization applied simultaneously to 5 pyramidal cells fails to increase their firing rate. 3) In modulated conditions, the same depolarization as in 2) yields reverberatory activity among the simulated group, while cells which were not depolarized are slightly inhibited (4). Reverberatory activity is terminated by a brief hyperpolarization applied to the 5 pyramidal cells.

The bistability in this model depended on a sufficient density of NMDA channels and a significant contribution of INap to the intrinsic excitability of the cell. The sustained firing of the cell could be terminated by a brief hyperpolarization (Fig 1A), or anincoming volley of inhibitory inputs (30 Hz average,200 ms). Modifying the strength of GABAergic andNMDA conductances could be used to modulate the frequency of discharge of the cell within the 20-45 Hz range. Lower firing frequencies could not be sustained for several seconds, and higher frequencies resulted in the inactivation of sodium channels and the termination of spiking in the depolarized state. This model showed further that this bistability was robust to significant variations in synaptic parameters (up to 24% in frequency, and up to 19 % in conductances).

This model was then morphologically simplified and reduced to 7 compartments. Calcium currents were removed and the conductance of IM was increased to re-establish spike frequency adaptation. A generic 3 compartmental interneuron model was then built (Ina and IK only). A fully connected network of 10 pyramidal cells and 2 interneurons was then set up.
Synaptic connections included AMPA/NMDA and GABA-A synapses (one per pair, Fig 1B). Somatic noise current was added to all cells to yield a spontaneous firing rate of about 5-7 Hz in pyramidal cells, in nonmodulated conditions. D1 modulation was simulated as a 45% reduction of IM, a 10% increase in INap, and a 40% increase in NMDA conductances, in accordance with recent experimental findings [12,17]. D1 modulation resulted in an overall lowering of the spontaneous firing rate of all pyramidal cells in the network (Fig 1B1). Five pyramidal cells were then briefly depolarized. In the modulated condition, this depolarization yielded sustained reverberating firing (~20-30 Hz) among the chosen cells, while the other pyramidal cells were inhibited by feedback inhibition from the interneurons (Fig 1B3,4 downregulated to <1 Hz, all pyramidal cells fired 2-5 Hz before stimulation). In nonmodulated condition, the pyramidal cell activity returned to baseline firing (5-7 Hz) after a short transient (Fig 1B2). These simulation show that network bistability can support delayed-sustained activity in a manner very similar to what is observed experimentally [3,7].

We then used two simplified models to study the effect of dopamine modulation of pyramidal neurons on stability of firing and sharpness of tuning curves during delay period activity in a large network. In the first model, we used a population of 1000 pyramidal neurons and 200 interneurons with a synaptic connectivity of 10% and uniform synaptic weights. Interneurons were modeled as basic integrate-and-fire units that match the physiological behavior of fast-spiking interneurons and GABA-A transmission[4]. Synaptic transmission was modulated by changing the probability of release ("synaptic reliability", [13]), and including synaptic depression. We modeled synaptic depression with a time course of 200 ms and a percentage of reduction of 30%, and assumed it to be present at the output of both pyramidal cells and interneurons [4]. For pyramidal neurons we used an enhanced version of the integrate-and-fire neuron ("temporal integrate-and-fire neurons", [14]) which was fitted to experimental measurements of unmodulated, deep-layer pyramidal cells in rat prefrontal cortex (with spike frequency adaptation (SFA)) and D1 modulated cells (no SFA). Finally, we connected pyramidal cells with 80% fast glutamatergic synapses (AMPA) and 20% NMDA synapses, and used two variants for the NMDA receptor: without D1 receptor activation, and with D1 receptor activation in accordance to experimental data [12,17]. The global reduction of glutamatergic output was modeled by a uniform 50% reduction of synaptic weights. This reduction proved sufficient to counterbalance the increase in activity due to D1 modulation of intrinsic properties.

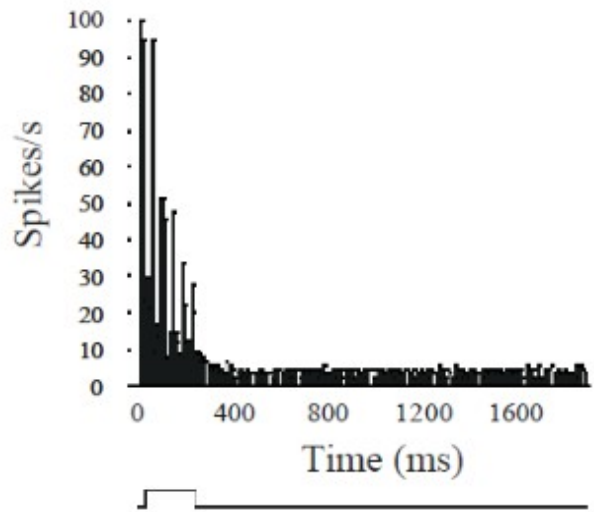
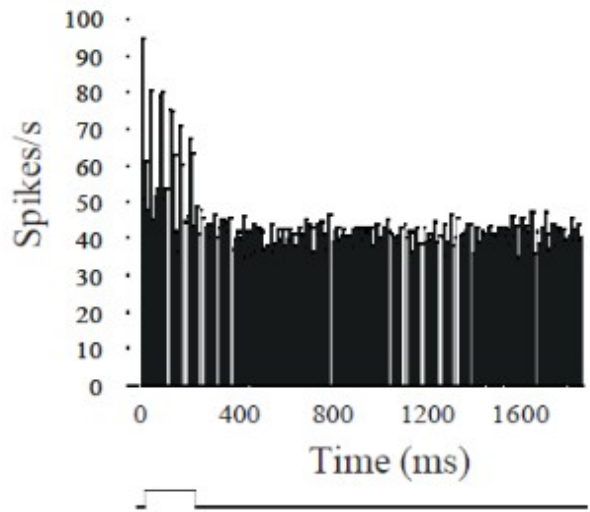

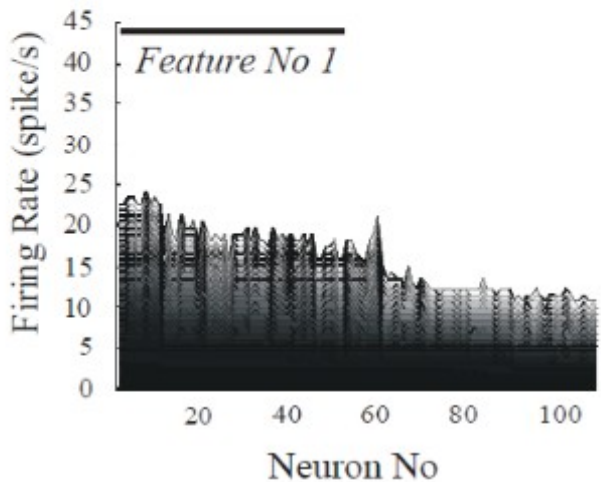
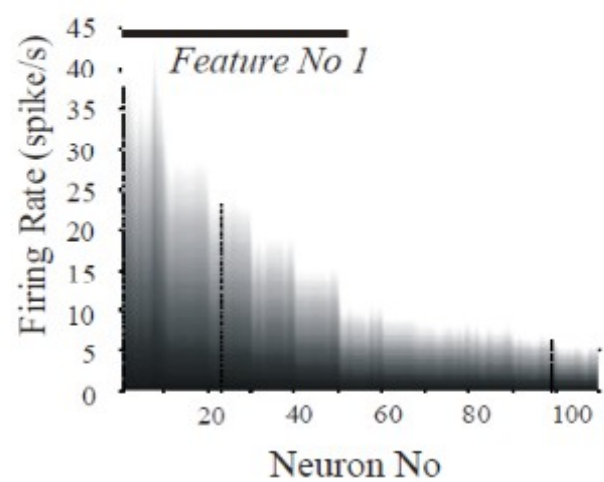

**Figure 2: A. Delay activity after a cue input of 250 ms in a network of 1000 pyramidal cells and 200 interneurons.** Right: nonmodulated neurons, mean firing rate 5Hz (background activity). Left: D1 receptor modulated neurons, mean firing rate 45 Hz. **B. Tuning curve in a network of 500 pyramidal cells and 100 interneurons.** Five features were stored in the network, so that each feature involved about 60 pyramidal cells. The first feature (coded by the first 60 pyramidal cells) was presented to the network. Left: maximal firing rate of the first 110 neurons (bin width 10ms) in unmodulated condition. The tuning curve is broad and shallow (neurons 1-60). Right: Maximal firing rate of the same neurons in modulated conditions. The tuning curve is sharp, and signal-to-noise is increased.

We then compared a set of neurons without D1 receptor activation and a set of neurons with D1 receptor activation. When both groups of neurons are unconnected and each is given a brief input (250ms) comparable to cue activity, we can see the difference in the stability and duration of the signal. In the D1 modulated group of neurons, activity is kept up for at least 2 s and results in high frequency interactions between neurons (40-60 Hz) (Fig. 2A right). In the control condition (no D1 modulation) an input signal results in a strong response that gradually fades within 100-800ms, depending on the strengths of synaptic connections (Fig. 2A left). Changing synaptic reliability (0.90.1) as a source of internal noise has little effect on the D1modulated network up to 0.5, but results in a considerable disturbance of processing in the control network. Looking at the contribution of individual parameters, we found that the modification of temporal dependence of neuronal processing due to the suppression of SFA and synaptic depression is sufficient to keep up delay activity, and NMDA enhancement can be supplanted by increasing synaptic strengths for AMPA or NMDA receptors.
However, we found that neurons fire less continuously and less synchronously without NMDA enhancement.

For the second model, we built a network of 500 pyramidal cells and 100 interneurons. We manually tuned the synaptic weights to achieve an attractor-like representation for five features and we then computed their tuning curves ("memory fields"). Each feature involved a population of 60 pyramidal cells. There have been indications that the shape of tuning curves in prefrontal cortex is influenced by D1receptor modulation [8,16] and it can be understood as an effect of selective 'attentional' enhancement for currently active features [9]. We looked at this question using our model of D1 activation on pyramidal neurons.
D1 modulation alone, by enabling higher frequencies for strongly activated neurons without affecting weakly activated neurons, increased the sharpness of tuning curves (Fig. 2B right, compared to left). Through feedback inhibition, the firing rates of neurons participating in the coding of other features were actually decreased, in effect increasing signal-to-noise ratio. The effect of dopamine modulation on locally tuned interneurons [7] probably raises inhibition by depolarizing the cells, and is expected to synergistically contribute to the sharpening of tuning curves.

## *Conclusion*
Working memory tasks require that memory items are temporarily stored by a population of cells. Experimental findings showed that prefrontal cortex cells elevate their firing during delay period activity if the memory item matches the cell preference. This elevation of firing rate depends on the co-activation of D1 dopamine receptors. Unlike short-term or longterm memory, working memory is a transient and flexible phenomenon that is activated and reset within a few hundreds milliseconds. This suggests that longterm synaptic plasticity such as LTP or LTD may not support this form of memory, and that the storing of memory reside in the dynamical binding of spatially specific populations of neurons whose 'memory field' matches the sensory input to store. We presented here a mechanism by which a specific subset of cells can be activated for several seconds following the brief presentation of a stimulus. This mechanism relies on the network bistability introduced by the effects of D1-like receptor activation of intrinsic and synaptic properties. In this regime, a subpopulation of cells is able to maintain an elevated rate of firing through reverberatory activity that is established and reset in few hundreds of milliseconds. Detailed and simplified multicompartmental techniques have been used to implement this mechanism, and show its feasibility.

Our results show that spatially specific patterns of activity can be robustly created and reset in a fully symmetric network (all synaptic weights equal), and that the D1 modulation of NMDA and INap currents is crucial for its implementation. A simplified model consisting of integrate-and-fire neurons further showed that, if synaptic weights are biased to store classical attractors, D1 modulation was able to sharpen the tuning curve of a specific attractor, and increase signal-to-noise ratio, through feedback inhibition. This finding links the subcellular mechanisms of dopamine neuromodulations to physiological measurements made in behaving animals during working memory tasks, and may serve as a testbed for further experimental studies on the role of dopamine in working memory task in prefrontal cortex and attentional enhancement.